    \documentclass[a4paper,oneside,Palatino]{article}
    \usepackage{latexsym}
    \usepackage{cite}
    \usepackage[dvips]{color}
    \usepackage[dvips]{graphicx}

\def\p{\partial}

\def\cF{\cal{F}}
\def\rN{\rm{N}}
\def\P{\Pi}
\catcode`@=11

\@addtoreset{equation}{section}
\catcode`@=12

\title{\centerline{\small MMV/No/08-10-15 \hfill arXiv:----.-----}\bigskip
\bf Photon \& Axion Oscillation In a Magnetized Medium: A Covariant Treatment}

\author{\bf Avijit K. Ganguly$^a$,
Pankaj Jain$^b$, Subhayan Mandal$^b$\\
\normalsize
$^a$ Dept. Of Physics, MMV, Beneras Hindu University, Varanasi-221005.\\
\normalsize
$^b$ Physics Department, Indian Institute Of Technology, Kanpur-208016,
India\\
\normalsize
{e-mail addresses:
avijitk@hotmail.com,
pkjain@iik.ac.in, shabsslg@iitk.ac.in}\\
}

\date{October  2008}

\begin{document}

\maketitle

%%%%%%%%%%%%%%%%%%%%%%%%%%%%%%  Abstract   %%%%%%%%%%%%%%%%%%%%%%%%%%%%%%%%%
%\begin{abstract}
\begin{center}
\begin{abstract}
\noindent
Pseudoscalar particles, with almost zero mass and very weak coupling 
to the visible matter, arise in many extensions of the standard model of
particle physics. Their mixing with photons in the presence of an external
magnetic field leads to many interesting astrophysical and cosmological 
consequences. This mixing depends on the   
medium properties, the momentum of the photon and the background magnetic 
field. Here we give a general treatment of pseudoscalar-photon oscillations
in a background magnetic field, taking the Faraday term into account.
We give predictions valid in all regimes, under the assumption that the
frequency of the wave is much higher than the plasma frequency of the
medium. At sufficiently high frequencies, the Faraday effect is negligible
and we reproduce the standard pseudoscalar-photon mixing phenomenon. However
at low frequencies, where Faraday effect is important, the mixing formulae
are considerably modified. We explicitly compute the contribution 
due to the longitudinal
mode of the photon and show that it is negligible.  
\end{abstract}
\end{center}
\noindent
%\end{center}
%\end{abstract}
%%%%%%%%%%%%%%%%%%%%%%%%%%%%% End Abstract %%%%%%%%%%%%%%%%%%%%%%%%%%%%%%%%%

%\end{titlepage}
%\newpage

\renewcommand{\thefootnote}{(\arabic{footnote})}

%%%%%%%%%%%%%%%%%%%%%%%%%%%%%%%%%%%%%%%%%%%%%%%%%%%%%%
%%%%%%%%%%%%%%%%%%%%%%%%%%%%%%%%%%%%%%%%%%%%%%%%%%%%%%
\section{Introduction}
\label{sec:intro}
\setcounter{equation}{0}
\setcounter{footnote}{0}
%%%%%%%%%%%%%%%%%%%%%%%%%%%%%%%%%%%%%%%%%%%%%%%%%%%%%%
%%%%%%%%%%%%%%%%%%%%%%%%%%%%%%%%%%%%%%%%%%%%%%%%%%%%%%
%%%%%%%%%%%%%%%%%%%%%%%%%%%%%%%%%%%%%%%%%%%%%%%%%%%%%%%%%%%%%%%

The standard model of particle physics suggests the existence of a very light,
weakly coupled particle, called the axion. It arises as a pseudo-Goldstone
boson of the broken Peccei Quinn (PQ) symmetry in a generalization of
the standard model \cite{PQ,Weinberg,McKay,kim1,Dine,McKay2,Kim}. 
Similar particles are also predicted by supergravity \cite{supergravity}
and superstrings theory \cite{Sen,Das}. 
Such a pseudoscalar particle has an effective 
coupling to two photons.
As a consequence, in an external magnetic field, 
an axion can oscillate to a photon and vice versa 
\cite{Karl,PS83,Maiani,RS88,Bradley,CG94,sudeep,Ganguly}.
The mixing can change both
the intensity and the state of polarization of the photons. 
Since the mixing is dependent on the frequency of the wave,
it also leads to a change in the spectrum of electromagnetic radiation
\cite{Ostman,Lai,Hooper,Hochmuth,Chelouche}.
This phenomenon has been used for laboratory and astrophysical searches 
for such particles, leading to stringent limits \cite{PDG06,Rosenberg,Brockway,astro,CAST,jaeckel,Robilliard,zavattini,Rubbia,Raffelt,Mohanty}.
Furthermore, the astrophysical and cosmological consequences of this mixing
have been studied extensively in the literature \cite{Raffelt,Mohanty,Csaki,PJ,sroy,MirizziCMB,MirizziTEV,Song,Gnedin,Payez,Piotrovich,Agarwal}. 
The mixing increases the transparency of the intergalactic and galactic medium
to propagation of high energy photons due to the very weak coupling of
pseudoscalars to visible matter \cite{Csaki1,Fairbairn,Angelis}.

Inside a  medium, there exists an additional loop
induced, axion-photon vertex, providing an extra contribution to
the usual axion-photon coupling in vacuum. In some kinematical limit
this contribution has  medium,  momentum  and  magnetic  field
dependence. In vacuum one of the two transverse modes of photon
gets coupled to an axion. In a medium the photon acquires
an extra degree of freedom (i.e., the longitudinal mode) and
as a consequence axion also couples to this additional degree of freedom.
In a magnetized medium, the two transverse degrees
of freedom are usually coupled due to the Faraday effect. Hence
in the presence of axions the two transverse modes 
would also get coupled to the 
longitudinal degree of freedom. 

In this paper we study the
axion-photon evolution, taking the Faraday term into account.
Our objectives in this study are two-folds. First we give a general
treatment of pseudoscalar-photon mixing in a medium. This allows us
to compute the next to leading order temperature dependent corrections. 
Next we give general solutions to the oscillation problem, taking into
account the Faraday effect. At very large frequencies we expect that the
Faraday effect would be negligible and the standard treatment of pseudoscala-photon mixing would apply. However at low frequencies the standard pseudoscalar-photon mixing result may deviate
significantly due to the presence of near degeneracies in the mixing 
matrix. In several laboratory and astrophysical situations this regime
may be applicable
and hence our treatment would be useful in 
such cases. In the present paper we restrict ourselves to providing a general 
treatment and do not address the issue of applications to laboratory 
experiments or to astrophysics.

%%%%%%%%%%%%%%%%%%%%%
\subsection{Interactions and the Polarization Tensor in a Medium}
%%%%%%%%%%%%%%%%%%%%%
The classical Lagrangian of a free electromagnetic field, including the gauge
fixing term is given by,
\begin{equation}
{\cal{L}}=-\frac{1}{4} {\rm{F}}_{\mu\nu}(x){\rm{F}}^{\mu\nu}(x) + \frac{1}{2\zeta}
\left(\partial_{\alpha} {\rm{A}}^{\alpha}(x)\right)^2 + j^{ext}.{\rm{A}}\,\,\,.
\label{lag}
\end{equation}
In Eq.(\ref{lag}), $\rm{F}^{\mu\nu}(x)=\left(\p^{\mu}\rm{A}^{\nu}(x)-\p^{\nu}\rm{A}^{\mu}(x)
\right )$, $\zeta$ is the gauge fixing parameter, $j_\mu^{\rm ext}$ is the 
external current and A(x)'s are the vector potential.
For the sake of simplicity, we use the Feynman gauge and set
$\zeta=1$.

As one takes quantum corrections into account, the quadratic part of the tree level Lagrangian
gets modified because of quantum corrections coming from terms proportional to the vacuum
polarization tensor $\Pi_{\mu\nu}$. The resulting Lagrangian, in the momentum space, is
\begin{eqnarray}
{\cal{L}}=\frac{1}{2} \left[ -k^2 {\tilde{g}}_{\mu\nu} + {{\rm{\Pi}}_{\mu\nu}(k) } \right]
{\rm{A}}^{\mu}(k){\rm{A}}^{\nu}(-k) - j^{ext}_\mu{\rm{A}}^{\mu}(k)
+ {{\cal{L}}_{G}}.
\label{mom-lag}
\end{eqnarray}
In Eq. (\ref{mom-lag}) above, $ {\cal{L}}_{G}$ corresponds to the gauge fixing term and
$\tilde{g}_{\mu\nu}=(g_{\mu\nu} -  \frac{k_{\mu}k_{\nu}}{k^2})$. In presence of a medium of
finite density and temperature, the polarization tensor get corrections from the matter
and temperature dependent parts. The resulting equation of motion for such a 
system can be written as,

\begin{eqnarray}
 \left[ -k^2 {\tilde{g}}_{\mu\nu} + \Pi_{\mu\nu}(k) \right] A^{\nu}(k)= j^{ext}_{\mu} ,
\label{eom1}
\end{eqnarray}
where we have not retained the pieces coming from the gauge fixing term 
(i.e. terms proportional to $\frac{k_{\mu}k_{\nu}}{k^2}$) since they can be shown to vanish. In a
medium composed of electrons, the polarization tensor, $\Pi_{\mu\nu}(k)$, can be written
in terms of scalar form factors and tensors composed out of the four vectors available for
the system, i.e., the medium four velocity $u^{\mu}=(1,0,0,0)$ and $k^{\mu}=({\omega, {\vec{k}}})$.
They are expressed as follows\cite{emftft},
\begin{eqnarray}
\Pi_{\mu\nu}(k) &=& \Pi_{T}R_{\mu\nu}+\Pi_{L} Q_{\mu\nu}\,, \nonumber \\
\mbox{where}\,\,&&\,\,
\left\{
\begin{array}{cc}
\!\!\!\!\!\!\!\!\!\!\!\!\!\! Q_{\mu\nu}= \frac{{\tilde{u}_{\mu}}{\tilde{u_{\nu}}}}{{{\tilde{u}^2}}}
\\
R_{\mu\nu}=\tilde{g}_{\mu\nu} -Q_{\mu\nu}\,, 
\end{array}
\right.
\label{pit1}
\end{eqnarray}
in absence of any external field. The vector $\tilde{u}_{\mu}$ is given by
$\tilde{u}_{\mu}=\tilde{g}_{\mu\nu}u^{\nu}$. The scalar form factor $\Pi_{L}(k)$,
corresponding to the longitudinal degree of freedom is given by,
\begin{eqnarray}
\Pi_{L}(k)&=&-\frac{k^2}{|{\vec{k}}^2|} \Pi_{\mu\nu}(k)u^{\mu}u^{\nu},
\mbox{~~~~where,~~~~}
u^{\mu}u^{\nu} \Pi_{\mu\nu}(k) =\omega^2_{p}
\left( \frac{|{\vec{k}}|^2}{\omega^2} +   3\frac{|{\vec{k}}|^4}{\omega^4}\frac{T}{m} \right)
%\nonumber \\
\end{eqnarray}
Similarly the transverse form factor $\Pi_{T}$ is given by the following expression:
\begin{eqnarray}
\Pi_{T}(k)&=& R^{\mu\nu}\Pi_{\mu\nu}(k)
\mbox{~~~~~~~~~~~~~~~~~~~and,~~~~}
{\rm{R}}^{\mu\nu} \Pi_{\mu\nu}(k) =\omega^2_{p}
\left( 1+  \frac{|{\vec{k}}|^2}{\omega^2}\frac{T}{m} \right).
\label{displ}
\end{eqnarray}
In the expressions above $\omega_p$ denotes the plasma frequency. In the classical limit,
to leading order in $\frac{T}{m}$, it is given by:
\begin{eqnarray}
\omega_p=\sqrt{\frac{4\pi\alpha n_e}{m}\left(1 - \frac{5T}{2m}\right)   }
\label{pf}
\end{eqnarray}
In Eq.~(\ref{pf}) $n_e$ is the number density of electrons.

We next express the form factors in terms of the dielectric constants.
Denoting, $F_{\mu\nu}=-i(k_{\mu}A_{\nu} -k_{\nu}A_{\mu})$, so that ${\vec{E}}=i\omega{\vec{A}}
-i\vec{k}A^0$ and $\vec{B} =i\vec{k}\times \vec{A}$, one can further define longitudinal
and transverse parts of the electric field as, $\vec{E_{l}}=\hat{k}(\hat{k}.\vec{E})$,  $E_{t}=
\vec{E} -\vec{E}_{l}$. Using the Kubo formula\footnote{~~~~i.e. $ j^{ind}_{\mu}(k) = -\Pi_{\mu\nu}(k)
A^{\nu}(k).$} for linear response analysis, the induced current,
$j^{ind}(k)$, can be written as,

\begin{eqnarray}
\vec{j}^{ind}=i\omega\left[ (1- \epsilon_l)\vec{E}_l+ (1-\epsilon_t)\vec{E}_t \right]
\end{eqnarray}
where longitudinal dielectric function $\epsilon_l$ and transverse dielectric function
$\epsilon_t$ are given as follows,

\begin{eqnarray}
\left(1 - \epsilon_l\right)= \frac{\Pi_L}{k^2} \nonumber \\
\left(1 - \epsilon_t\right)= \frac{\Pi_T}{\omega^2}
\end{eqnarray}

Using the relations given by Eq. (\ref{displ}) for $\Pi_T(k)$ and $\Pi_L(k)$, one can further
express the dielectric functions in terms of $\omega$ and $\vec{k}$. \\

%%
%%%%%%%%%%%%%%%%%%%%%%%%%%%%%%%%%%%%%%%%%%
\section{Inclusion of the Faraday Term}
%%%%%%%%%%%%%%%%%%%%%%%%%%%%%%%%%%%%%%%%%
\subsection{Exact Faraday Term}
%%%%%%%%%%%%%%%%%%%%%%%%%%%%%%%%%%%%%%%%%
The contribution to the vacuum polarization
tensor which is odd in $\cal B$ has been estimated using real time finite temperature field theory in Refs.  
\cite{faraday,jose}. It can be expressed as,  
\begin{eqnarray}
\Pi^{F}_{\lambda\rho}(k) &=& 4ie^2 \int \frac{d^4p}{(2\pi)^4} \eta_-(p)
\int_{-\infty}^\infty ds \; e^{\Phi(p,s)}
\int_0^\infty ds' \; e^{\Phi(p',s')} \Big[ R^{(1)}_{\lambda\rho} +
R^{(2a)}_{\lambda\rho} \Big] \nonumber\\
&=& 4ie^2 \varepsilon_{\lambda\rho\alpha_\parallel\beta} k^\beta
\int \frac{d^4p}{(2\pi)^4} \eta_-(p)
\int_{-\infty}^\infty ds \; e^{\Phi(p,s)}
\int_0^\infty ds' \; e^{\Phi(p',s')} \nonumber\\*
&\times & \Bigg[
p^{\widetilde\alpha_\parallel} \tan e{\cal B}s +
p'^{\widetilde\alpha_\parallel} \tan e{\cal B}s'
- {\tan e{\cal B}s \; \tan e{\cal B}s' \over \tan e{\cal B}(s+s')} \;
(p+p')^{\widetilde\alpha_\parallel} \Bigg] \,.
\label{finall}
\end{eqnarray}
\noindent
This is exact to all orders in $eB$. \\

\noindent
In Eq. (\ref{finall}) $\alpha_{\parallel}$ stands for 0 or the 3.
 Moreover appearence of $\alpha_{\parallel}$ and 
$\alpha_{\tilde{\parallel}}$ together in any product would mean: if $\alpha_{\parallel}$ takes the value  
0, $\alpha_{\tilde{\parallel}}$ would take the value 3 and vice versa. 
Furthermore for thermal fermions, $\eta_F(p)$ is the distribution function,
\begin{eqnarray}
\eta_F(p) = \Theta(p\cdot u) f_F(p,\mu,\beta) 
+ \Theta(-p\cdot u) f_F(-p,-\mu,\beta) \,.
\label{eta}
\end{eqnarray}
Here, $\Theta$ is the step function, which takes the value
$+1$ for positive values of its argument and vanishes for
negative values of the argument, and $f_F$ denotes the
Fermi-Dirac distribution function,
\begin{eqnarray}
f_F(p,\mu,\beta) = {1\over e^{\beta(p\cdot u - \mu)} + 1} \,.
\end{eqnarray} \\
Lastly   $\phi(p,s)$ in the exponential stands for, 
\begin{eqnarray}
\Phi(p,s) = is \left( p_\parallel^2 - {\tan
(e{\cal B}s) \over e{\cal B}s} \, p_\perp^2 - m^2 \right) - \epsilon
|s| \,, 
\label{Phi}
\\
\end{eqnarray}
However performing the integrals in Eq. (\ref{finall})
and arriving at a compact form is difficult.
This integral, can be performed in the 
long wavelength limit. 
However in this case, the expression does not look gauge invariant unless the
explicit long wavelength limit is maintained.
%%%%%%%%%%%%%%%%%%%
\subsection{Equation of Motion with Faraday Contribution}
\label{ss2}
%%%%%%%%%%%%%%%%%%
\noindent
In presence of external mgnetic field, Eq. (\ref{eom1}) is further modified to
\cite{faraday,jose}
\begin{eqnarray}
 \left[ -k^2 {\tilde{g}}_{\mu\nu} + \Pi_{\mu\nu}(k) + \Pi^{p}_{\mu\nu}(k) \right]
A^{\nu}(k)= j^{ext}_{\mu}.
\label{eom2}
\end{eqnarray}
We define,
\begin{displaymath}
|K|= \left[ \sum^{3}_{i=1}k^{2}_{i}\right]^{\frac{1}{2}}.
\end{displaymath}
In the limit $|K| \to 0$, one can express,
$$\Pi^{p}_{\mu\nu}(k)=\Pi^p(k)i\epsilon_{\mu_{\perp}\nu\alpha\beta}
\frac{k^{\alpha}}{|K|}u^{\beta}= \Pi^p(k)P_{\mu\nu},$$ with
$$P_{\mu\nu} = i\epsilon_{\mu_{\perp}\nu\alpha\beta}\frac{k^{\alpha}}{|K|}u^{\beta}. $$
The limit $ |K| \to 0$ should be taken in such a way that
\begin{displaymath}
{\cal{L}}t_{|K|\to 0} \left( \frac{k^{i}}{|K|} \right) \to 1.
\end{displaymath}
\noindent
The scalar form factor associated with the Faraday rotation term is given by
\cite{jose}
\begin{eqnarray}
 \Pi^p(k) = \frac{\omega \omega_{B} \omega^2_p}{\omega^2-\omega^2_{B}}, \mbox{~~where~~~}
\omega_{B}=\frac{eB}{m}.
\label{faraday}
\end{eqnarray}
%
%%%%%%%%%%%%%%%%%% Subsection Axion Electrodynamics  %%%%%%%%%%%%%%%%%%%%
%%%%%%%%%%%%%%%%%%%%%%%%%%%%%%%%%%%%%%%%%%%%%%%%%%%%%%%%%%%%%%%%%%%%%%%%%
\subsection{Axion Electrodynamics}
%%%%%%%%%%%%%%%%%%%%%%%%%%%%%%%%%%%%%%%%%%%%%%%%%%%%%%%%%%%%%%%%%%%%%%%%%
%%
%
In this subsection we obtain an expression for the 
effective Lagrangian for axion-electrodynamics in a 
magnetized medium.  
It is important
to note here that the axion contribution to the effective Lagrangian in a magnetized medium
does get modified due to the presence of medium and magnetic field.
However all the modifications
can be included by redefining the axion-photon coupling constant
\cite{Ganguly}.
The structure of the interaction Lagrangian remains identical to that at
tree level. In light
of this observation, we work with the tree level axion-photon Lagrangian in a magnetic field.
\noindent
In momentum space this effective Lagrangian is given by:,
\begin{eqnarray}
{\cal{ L}}= \frac{1}{2}\left[-A_{\mu} k^2\tilde{g}^{\mu\nu}
A_{\nu} + A_{\mu} \tilde{\Pi}^{\mu\nu}A_{\nu} +
i\frac{{\tilde{{\cF}}}^{\mu\nu}k_{\mu}A_{\nu} a}{2M_a} -a(k^2- m^2)a
\right]. \label{apl}
\end{eqnarray}\\

\noindent
In the expression above, $\tilde{\cF}^{\mu\nu}$ corresponds to the field strength tensor of the external field,$A_{\mu}$ is the vector potential for the 
photon, $a$ is the axion field, $1/M_a$ is the pseudoscalar-photon coupling and $ \tilde{\Pi}^{\mu\nu} $ is the
polarization tensor in matter along with the Faraday contribution,
\begin{equation}
\tilde{\Pi}^{\mu\nu}(k) = \Pi_{T}(k)R^{\mu\nu}
+\Pi_{L}(k)Q^{\mu\nu}(k)+\Pi_{p}(k)P^{\mu\nu} \label{pi-exp}\,.
\end{equation}
\noindent
The equations of motion for the Lagrangian given by Eq. (\ref{apl}) are the following:
\begin{eqnarray}
\left( -k^2\tilde{g}_{\alpha\nu}+\tilde{\Pi}_{\alpha\nu} (k) \right)A^{\nu} (k)=-i\frac{k^{\mu}
\tilde{\cF}_{\mu\alpha}a}{2M_a}
\label{eom-A}
\end{eqnarray}
\begin{eqnarray}
\left( k^2 -m^2 \right)a = i\frac{b^{(2)}_\mu A^{\mu}(k)}{2M_a}
\label{eom-ax}
\end{eqnarray}
%
%%%%%%%%%%%%%%%%%%%%%%%%%%%%55 New definition %%%%%%%%%%%%%%%%%%%%%%%%%
\def\cD{{\cal{D}}}
%%%%%%%%%%%%%%%%%%%%%%%%%%%%%%%%%%%%%%%%%%%%%%%%%%%%%%%%%%%%%%%%%%%%%%%
%
From now on we would denote $b^{(2)}_{\alpha}= k^{\mu}\tilde{\cF}_{\mu\alpha} $.

%%%%%%%%%%%%%%%%%%%%%5 Subsection %%%%%%%%%%%%%%%%%%%%%%%%%%%%%%
\subsection{Expanding $A^{\mu}(k)$ in Orthogonal Basis.}
%%%%%%%%%%%%%%%%%%%%%%%%%%%%%%%%%%%%%%%%%%%%%%%%%%%%%%%%%%%%%%%%

\noindent
In this subsection we construct a system of orthonormal basis vectors
out of the vectors available to us. Recall that the available vectors 
we have at our disposal are, $u^{\mu}$ the 4 velocity of the medium and
$k^{\mu}$ the external
momentum of the photon. Using these two and $\cF^{\mu\nu}$ we can further construct
two more vectors, 
\begin{equation}
b^{(1)\nu}=k_{\mu}\cF^{\mu\nu}\,, 
\label{bv1} 
\end{equation}
and
\begin{equation}
b^{(2)\nu}=k_{\mu}{\tilde{\cF}}^{\mu\nu}\,. 
\label{bv2} 
\end{equation}
We also define,
\begin{equation}
\tilde{u}^{\nu}=\left(g^{\mu\nu}-\frac{k^{\mu}k^{\nu}}{k^2}  \right)u_{\mu}\,.
\nonumber
\end{equation}
We note that, the vectors $b^{(1)}$, $b^{(2)}$ and $\tilde{u}$ are all
orthogonal to $k$. Now we can construct
another vector, $I^{\nu}$, such that it is orthogonal to both $\tilde{u}$ and $b^{(2)}$,
\begin{eqnarray}
I^{\nu}=\left( b^{(2)\nu} - \frac{(\tilde{u}^{\mu} b^{(2)}_\mu)}{\tilde{u}^{2}}
\tilde{u}^{\nu} \right)\,.
\label{bvv}
\end{eqnarray}
It is easy to verify that the vectors, $I^{\nu}$
$\tilde{u}^{\nu}$, $k^{\nu}$, $b^{(1)\nu}$ are mutually orthogonal
to each other.
Therefore we can express the gauge potential as their 
linear combination,
\begin{eqnarray}
A_{\alpha}(k)= A_{1}(k)\rm{N}_1 b^{(1)}_{\alpha}+ A_{2}(k) \rm{N}_2I_{\alpha}+ A_{L}(k)
\rm{N}_L\tilde{u}_{\alpha} + k_{\alpha}\rm{N}_{\parallel}A_{||}(k).
\label{vec-A}
\end{eqnarray}
The component, $A_{||}(k)$ can be set to zero, since it is associated with
the gauge degrees of freedom. In Eq. (\ref{vec-A}) the
$\rm{N}_{i}$'s are normalization constants for the
corresponding basis vectors.  
We can easily see that the normalization constants
are given such to be,
\begin{equation}
N_1 = {1\over\sqrt{-b^{(1)}_\mu b^{(1)\mu}}}
\end{equation}
\begin{equation}
N_2 = {1\over\sqrt{-I_\mu I^{\mu}}}
\end{equation}
\begin{equation}
N_L = {1\over\sqrt{-\tilde{u}_\mu \tilde{u}^{\mu}}}
\end{equation}
The negative sign under the square root is for maintenance of the
reality of the normalization constants. 
These constants take the values,

$$N_1 = {1\over B_zK_\perp} $$
$$N_2 = {|\vec{K}| \over \omega K_\perp B_z} $$
$$N_L = {K \over |\vec{K}|}$$

%
%%%%%%%%%%%%%%%%%%%%%%%%%%%%%%%%%%%%%%%%%%%%%%%%%%%%%%%%%%%%%%%%%%%%%%%%%%%%%%%%%%%%%%
%%%%%%%%%%%%%%%%%%%%%%%%%%%%%%%%%%% Section %%%%%%%%%%%%%%%%%%%%%%%%%%%%%%%%%%%%%%%%%%
%%%%%%%%%%%%%%%%%%%%%%%%%%%%%%%%%%%%%%%%%%%%%%%%%%%%%%%%%%%%%%%%%%%%%%%%%%%%%%%%%%%%%%
\section{Equation of Motion in terms of the Form Factors}
%%%%%%%%%%%%%%%%%%%%%%%%%%%%%%%%%%%%%%%%%%%%%%%%%%%%%%%%%%%%%%%%%%%%%%%%%%%%%%%%%%%%%%
%%%%%%%%%%%%%%%%%%%%%%%%%%%%%%%%%%%%%%%%%%%%%%%%%%%%%%%%%%%%%%%%%%%%%%%%%%%%%%%%%%%%%%
%
\noindent
The equation of motion for photon, Eq. (\ref{eom-A}), can be written in
a expanded form, 
\begin{eqnarray}
\left[ k^2g_{\mu\nu}-\Pi_{T}(k)R_{\mu\nu} -\Pi_{L}(k)Q_{\mu\nu}(k)- \Pi_{p}(k)P_{\mu\nu} \right]
A^{\nu}(k)= i\frac{b^{(2)}_\mu a}{2M_a}
\label{eom-A2}
\end{eqnarray}
where we have used Eq. (\ref{pi-exp}).
In order to arrive at the equations of motions in terms of the
form factors, one needs to substitute Eq. (\ref{vec-A}) in Eq.
(\ref{eom-A2}) and project out the different components of this equation. 
Multiplication from left by the normalized basis vectors, $N_1\times
b^{(1)}_{\mu}$, $N_2\times I_{\mu}$ and $N_L\times \tilde{u}_{\mu}$ 
leads to the equations, 
\begin{eqnarray}
-(k^2 - \P_T(k))A_{2}(k) + i\P_{p}N_1N_2\left[
\epsilon_{\mu_{\perp}\nu_{\perp}30}b^{(1)\nu} I^{\mu}
\right]{\rN_1}A_{1}(k)\!\!\!\!&=&\!\!\!\! \frac{
\left(i N_2b^{(2)}_{\mu}I^{\mu} \right)a}{2M_a}\,,
\nonumber \\
(k^2 -\P_T(k))A_{1}(k)+
i\P_{p}N_1N_2\left[\epsilon_{\mu_{\perp}\nu_{\perp}30}b^{(1)\mu}
I^{\nu} \right]A_{2}(k)\!\!\!\!\! &=&\!\!\!\! 0\,, \nonumber \\
\left(k^2 - \Pi_L \right)A_L(k)\!\!\!\!&=&\!\!\!\!
\frac{i N_L\left(b^{(2)}_\mu\tilde{u}^{\mu}
\right)a}{2M_a}  
\label{eom-fs}
\end{eqnarray}
respectively. The equation of motion for the pseudoscalar field can be 
expressed as,
\begin{equation}
\left[\frac{\left( i b^{(2)}_{\mu}I^{\mu}\! \right)}{2M_a}{\rN}_2 A_{2}(k)
+ \frac{\left( i  b^{(2)}_{\mu} \tilde{u}^{\mu} \!\right)}{2M_a}{\rN}_L A_{L}(k) \right]
 = \left(k^2 -m^2 \right)a.
\label{eom-fs1}
\end{equation}
This completes the closed set of equations involving the axion and the photon vector
potential. It is easy to see that, if one sets the axion field to be equalto zero,
one recovers the usual Maxwell Equations modified by the Faraday pieces.
%%%%%%%%%%%%%%%%%%%%%%%%%%%%%%%%%%%%%%%%%%%%%%%%%%%%%%%%%%%%%%%%%%%%%%%%%%%%%%%%%%%%%%
%%%%%%%%%%%%%%%%%%%%%%%%%%%%%%%%%% Section %%%%%%%%%%%%%%%%%%%%%%%%%%%%%%%%%%%%%%%%%%%
%%%%%%%%%%%%%%%%%%%%%%%%%%%%%%%%%%%%%%%%%%%%%%%%%%%%%%%%%%%%%%%%%%%%%%%%%%%%%%%%%%%%%%
\section{ Mixing Matrix }
%%%%%%%%%%%%%%%%%%%%%%%%%%%%%%%%%%%%%%%%%%%%%%%%%%%%%%%%%%%%%%%%%%%%%%%%%%%%%%%%%%%%%%
%%%%%%%%%%%%%%%%%%%%%%%%%%%%%%%%%%%%%%%%%%%%%%%%%%%%%%%%%%%%%%%%%%%%%%%%%%%%%%%%%%%%%%
One can observe from Eqs. (\ref{eom-fs}), (\ref{eom-fs1}) that, the coupling to axions mixes
the longitudinal component $A_{L}$ to the transverse components. So even if $A_{L}$
is zero to begin with, the same can be generated through coupling through the pseudoscalar
field a. However the coupling is suppressed by the PQ symmetry breaking scale. 
One needs to study these equations carefully to determine the effect of
pseudoscalar-photon coupling and magnetized medium 
on the Electro-Magnetic (EM) vector potentials.

We are interested in a quasi-monochromative wave solution. We shall assume
that the wave propagates in the $z$ direction and express the solution in the
form
\begin{equation}
\phi_i(t,z) = e^{-i\omega t}\phi_i(0,z)
\end{equation}
where $\phi_i$ may represent the pseudoscalar field or any of the
components of the electromagnetic wave.
We work in the eikonal limit, where $w\approx k$ is the largest energy scale. 
We may now express 
Eqs. (\ref{eom-fs}), (\ref{eom-fs1}) in real space in the matrix form
\begin{equation}
\left[(\omega^2 + \partial_z^2){\bf I} - M\right]
\left( \begin{array}{c}
A_{1}(k)   \\
A_{2}(k)   \\
A_{L}(k)  \\
a (k)  \\
\end{array} \right)=0.
\end{equation}
where ${\bf I}$ is a $4\times 4$ identity matrix and the mixing matrix,
\begin{equation}
M = \left( \begin{array}{cccc}
 \Pi_T & - i\P_p\epsilon_{\mu_{\perp}\nu_{\perp}30}b^{(1)\mu}I^{\nu} & 0 & 0 \\
i\P_p\epsilon_{\mu_{\perp}\nu_{\perp}30}b^{(1)\nu}I^{\mu}   & +
\P_T
& 0 & -i\frac{N_2b^{(2)}_\mu I^{\mu}}{2M_a} \\
0 & 0 &  {\P_L}     &   -i\frac{N_Lb^{(2)}_\mu \tilde{u}^{\mu}}{2M_a}\\
0 & i \frac{N_2b^{(2)}_\mu I^{\mu}}{2M_a} &  i
\frac{N_Lb^{(2)}_\mu \tilde{u}^{\mu}}{2M_a}  & m^2_a
\end{array} \right)
\label{mf-1}
\end{equation}
In order to find out axion-photon oscillation, we need to diagonalize 
the matrix given in Eq. (\ref{mf-1}). Although it can be diagonalized
exactly, the resulting formulae are too cumbersome to be directly useful. 
We make some simplifying assumptions to make the problem tractable.
The longitudinal component has been shown to contribute 
negligibly in Ref. \cite{sudeep1}. Hence at leading order 
we set $A_L(z) = 0$ and compute
its contribution perturbatively. 

Our final aim is to compute the Stokes parameters $I(z),\ Q(z),\ U(z),\ V(z)$
at some distance $z$, given the values of these parameters at the origin
$z=0$. For this purpose we may compute the density matrix
\begin{equation}
\rho(z) = \left(\matrix{<A_1 A_1^*> & <A_1 A_2^*> & <A_1 a^*>\cr
 <A_2 A_1^*>  & <A_2 A_2^*> & <A_2 a^*>\cr
 <a A_1^*> & <a A_2^*> & <a a^*>}\right)
\end{equation}
where the angular brackets $< >$ represent
ensemble averages.
We work in the Lorentz gauge $\Box{A}=0$ and hence from these matrix we can directly
compute the coherency matrix
\begin{equation}
\bar\rho(z) = \left(\matrix{<E_1 E_1^*> & <E_1 E_2^*>\cr
 <E_2 E_1^*>  & <E_2 E_2^*>}\right)
\end{equation}
where $E_i$ are the electric fields.

\section{Solutions}
We now solve the equations of motion, ignoring the longitudinal mode. 
We work in the low temperature limit and ignore the terms proportional 
to temperature. We may express the resulting mixing matrix $M$ 
as 
\begin{equation}
M = \left(\matrix{A & iF & 0\cr
           -iF & A & -iT\cr
	   0 & iT  & B}\right)\,,
\label{eq:matrix33}
\end{equation}
where $A = \omega_p^2$, $B=m_a^2$, $F = \omega \omega_B\omega_p^2\cos\theta
/(\omega^2-\omega_B^2)$, and $T=|\vec B|\omega\sin\theta/2M_a$. Here 
$\theta$ is the angle between the background magnetic field and the direction
of propagation. 
Throughout we shall assume that the background
magnetic field is independent of space and time. The component transverse 
to the direction of propagation is taken to point along the `$y$' or `2' 
axis.

We denote the eigenvalues of $M$ by $\lambda_i$. 
We define an alternate matrix $\bar M$,
\begin{equation}
{\bar M}=(M-AI)/(B-A) = \left(\matrix{0 & ix & 0\cr
           -ix & 0 & -iy\cr
	   0 & iy  & 1}\right)\,,
\label{eq:matrix331}
\end{equation}
where $I$ is an identity matrix, $x=F/(B-A)$ and $y=T/(B-A)$. The eigenvalues
$\bar\lambda_i$ of $\bar M$ are related to $\lambda_i$ by
\begin{equation}
\lambda_i = \bar\lambda_i(B-A) + A\,.
\label{eq:lambdabar}
\end{equation}
The eigenvectors of $\bar M$ (or $M$) may be expressed as
\begin{equation}
|\bar\lambda_i> = {1\over D_i}\left(\matrix{x(\bar\lambda_i-1)\cr -i\bar\lambda_i
(\bar\lambda_i-1)\cr y\bar\lambda_i}\right)\,,
\end{equation}
where $D_i^2 = 2x^2(\bar\lambda_i^2-1)^2 + y^2(2\bar\lambda_i^2-\bar\lambda_i)$.
The eigenvalues $\bar\lambda_i$ may be written as
\begin{eqnarray}
\bar\lambda_1 &=& 2 \sqrt{-Q}\cos\left({\theta+4\pi\over 3}\right) + 
{1\over 3}\,,\nonumber\\
\bar\lambda_2 &=& 2 \sqrt{-Q}\cos\left({\theta+2\pi\over 3}\right) + 
{1\over 3}\,,\nonumber\\
\bar\lambda_3 &=& 2 \sqrt{-Q}\cos\left({\theta\over 3}\right)+{1\over 3}\,,
\end{eqnarray}
where  
\begin{eqnarray}
\theta &=& \cos^{-1}\left({R\over \sqrt{-Q^3}}\right)\,,\nonumber\\
R &=& {1\over 54}(9y^2-18x^2+2)\,,\nonumber\\
Q &=& -{x^2+y^2\over 3} - {1\over 9}\,.
\end{eqnarray}
We point out that for a cubic equation we expect three real eigenvalues
if $Q^3+R^2<0$, which must of course be valid in our case. 

Using this we construct the unitary matrix $U$ to diagonalize $M$ and hence
solve the propagation equation. The final result for the density matrix after 
propagating a distance $z$ from the initial point is 
\begin{equation}
\rho(z) = UPU^\dagger\rho(0)UP^\dagger U^\dagger\,,
\end{equation}
where $P$ is the propagation matrix,
\begin{equation}
P= \left(\matrix{\exp(ik_1z) & 0 & 0\cr
           0 &\exp(ik_2 z) & 0\cr
	   0 & 0  & \exp(ik_3z) }\right)\,,
\label{eq:propmat}
\end{equation}
and $k_i= \sqrt{\omega^2-\lambda_i}\approx \omega-{\lambda_i\over 2\omega}$.

Although the problem is solvable exactly, we find that there exist parameter
ranges where the exact formulas are not very useful for numerical 
calculations. Furthermore in order to get some insight into the solution
it is useful to explore it analytically in some limiting cases. 
The problem is a little tricky due to near degeneracy of the two modes 
of the electromagnetic wave.

For astrophysical and cosmological applications we are usually interested
in the case where both the parameters $T$ and $F$ are very
small compared to $|A-B|$, i.e. $x<<1$ and $y<<1$. 
This is the case of weak magnetic field and pseudoscalar-photon coupling. 
In laboratory experiments the parameter $y$ may not be small compared to unity
if the plasma density is taken to be very small. We 
postpone a detailed discussion of laboratory experiments to a separate 
publication. For now we discuss the astrophysically interesting limit
of $x<<1$ and $y<<1$. 

We expand the eigenvalues in powers of $x$ and $y$. 
This expansion is a little messy since we find that we need to include
terms atleast up to order $y^4$. This is because in the limit $x\rightarrow 0$, 
these terms contribute at leading order.
The expression 
$R/\sqrt{-Q^3}\rightarrow 1$ in the limit $x \rightarrow 1$, $y \rightarrow 1$. 
Hence we may express
\begin{equation}
R/\sqrt{-Q^3} = 1-\alpha-\beta\,,
\end{equation}
where 
\begin{eqnarray}
\alpha &=& 27x^2/2 <<1\,,\nonumber\\
\beta &=& {27\over 8}y^4 - {27\times 17\over 8} x^4 - 54 x^2y^2<<1\,.
\end{eqnarray}
The $\cos^{-1}(1-\alpha-\beta)$ 
function has a non-analytic structure in the neighbourhood of unity. 
Expanding in powers of $x$ and $y$ we find \cite{stegun}
\begin{equation}
\theta = \cos^{-1} (1-\alpha-\beta) = {\sqrt{\alpha+\beta}\over \sqrt{2}}
\left[2+{\alpha+\beta\over 6} + {3(\alpha+\beta)^2\over 80} + ...\right]\,.
\end{equation}
The eigenvalues are approximately given by
\begin{eqnarray}
\bar\lambda_1 &=& -{1\over 2}(x^2+y^2)+{2\over 3\sqrt{6}}\sqrt{\alpha+\beta}
+ {3\over 8} (x^2+y^2)^2 + {1\over \sqrt{6}}(x^2+y^2)\sqrt{\alpha+\beta} 
\nonumber\\
&+&{1\over 27}(\alpha+\beta)+{\alpha\over 18}(x^2+y^2)\,,\nonumber\\
\bar\lambda_2 &=& -{1\over 2}(x^2+y^2)-{2\over 3\sqrt{6}}\sqrt{\alpha+\beta}
+ {3\over 8} (x^2+y^2)^2 - {1\over \sqrt{6}}(x^2+y^2)\sqrt{\alpha+\beta}
\nonumber\\
&+&{1\over 27}(\alpha+\beta)+{\alpha\over 18}(x^2+y^2)\,,\nonumber\\
\bar\lambda_3 &=& 1 + (x^2+y^2) - {2\over 27}(\alpha+\beta) - {3\over 4}
(x^2+y^2)^2 - {1\over 9}(\alpha+\beta)(x^2+y^2)\,.
\end{eqnarray}
Here we have displayed terms accurate to order $x^2, y^4$ and $x^2y^2$. 
We again emphasize that in the limit $x\rightarrow 0$, the dominant 
contribution comes from terms proportional to $y^4$. 
Hence we need expand up to order $y^4$. As we can see the expansion
is quite cumbersome and does not yield a simple analytic result. 
However it is suitable for numerical calculations in the limit the parameters
$x$ and $y$ are very small.
We next consider some limiting cases in which a simple analytic result can be obtained.

\subsection{Limit 1: $T<<F<<|A-B|$}
This is the regime where the Faraday effect is dominant and hence
 relevant at low frequencies. We expand the 
eigenvalues and eigenvectors in powers of $T$ and
obtain results accurate to order $T^2$. 
This expansion may also be obtained directly by using perturbation theory.
We may first exactly diagonalize the matrix $M$ in the limit $T=0$ and then
compute leading order corrections in $T$.
The resulting eigenvalues are
given by
\begin{eqnarray}
\lambda_1 &=& A-F+{T^2\over 2(A-F-B)}\nonumber\\
\lambda_2 &=& A+F+{T^2\over 2(A+F-B)}\nonumber\\
\lambda_3 &=& B-{T^2\over 2(A-F-B)} - {T^2\over 2(A+F-B)}
\end{eqnarray}
with the corresponding eigenvectors
\begin{eqnarray}
|\lambda_1> &=& {1\over \sqrt{2}} \left(\matrix{
1-{T^2\over 4(A-F-B)^2}+{T^2\over 4F(A-F-B)}\cr
i\left[1-{T^2\over 4(A-F-B)^2}-{T^2\over 4F(A-F-B)}\right]\cr
-{T\over A-F-B}} \right)\nonumber \\
|\lambda_2> &=& {1\over \sqrt{2}} \left(\matrix{
i\left[1-{T^2\over 4(A+F-B)^2}-{T^2\over 4F(A+F-B)}\right]\cr
1-{T^2\over 4(A+F-B)^2}+{T^2\over 4F(A+F-B)}\cr
i{T\over A+F-B}} \right)\nonumber \\
|\lambda_3> &=& {1\over \sqrt{2}} \left(\matrix{
{T\over 2(A-F-B)}-{T\over 2(A+F-B)}\cr
i{T\over 2(A-F-B)}+i{T\over 2(A+F-B)}\cr
1-{T^2\over 4(A-F-B)^2}-{T^2\over 4(A+F-B)^2}} \right)
\end{eqnarray}

In the limit under consideration we may expand the denominators in powers 
of $F/(A-B)$. Since $F>>T$, we compute the density matrix accurate to order
$T^2/[F(A-B)]$ and drop terms proportional to $T^2/(A-B)^2$. 
We also assume that all correlators involving the pseudoscalar field are
zero initially. The relevant elements of the density matrix are given
by
\begin{eqnarray}
\rho_{11}(Z) &=& {1\over 2}(\rho_{11}(0)+\rho_{22}(0))+{\cos (\Delta_{12}Z) 
\over 2}
(\rho_{11}(0)-\rho_{22}(0))\nonumber\\ 
&+& \left(\rho_{12}(0)\left[-{1\over 2}\sin(\Delta_{12} Z) + {iT^2\over 
4F(A-B)}(1-\cos(\Delta_{12} Z))\right] + c.c.\right)\nonumber\\
\rho_{22}(Z) &=& {1\over 2}(\rho_{11}(0)+\rho_{22}(0))-{\cos(\Delta_{12} 
Z)\over 2}
(\rho_{11}(0)-\rho_{22}(0)) \nonumber\\
&+& \left(\rho_{12}(0)\left[{1\over 2}\sin(\Delta_{12} Z) - {iT^2\over 
4F(A-B)}(1-\cos(\Delta_{12} Z))\right] + c.c.\right)\nonumber\\
\rho_{12}(Z) &=& {\sin(\Delta_{12} Z)\over 2}(\rho_{11}(0)-\rho_{22}(0)) 
+ {iT^2\over 4F(A-B)}(1- \cos(\Delta_{12} Z))
(\rho_{11}(0)+\rho_{22}(0))\nonumber\\
&+&{\rho_{12}(0)\over 2}\left[1+\cos(\Delta_{12} Z)- 
{iT^2\over F(A-B)}\sin(\Delta_{12} Z)
\right] + {\rho^*_{12}(0)\over 2}\left[-1+\cos(\Delta_{12} Z)\right]
\end{eqnarray}
where $\Delta_{12} = (\lambda_1-\lambda_2)/(2\omega)$.
We find the usual Faraday rotation along with additional contributions 
proportional to $T^2/[F(A-B)]$. 
If the initial beam is unpolarized or linearly polarized, 
pseudoscalar-photon will generate circular polarization. 
The dominant contribution arises if the initial beam has a non-zero real
part of $\rho_{12}$ or equivalently a non-zero Stokes parameter $U$. In
this case in the limit $\Delta_{12}Z<<1$, the circular polarization 
generated is proportional to $T^2Z/[\omega(A-B)]$. Hence this contribution
is proportional to $\omega$. 

\subsection{Limit 2:  $F/|A-B|<<T^2/|A-B|^2<<1$}
We next consider 
another limiting case where the Faraday rotation effect is negligible
and pseudoscalar-photon mixing dominates. This is the case where the leading
order contribution leads to the standard pseudoscalar-photon mixing. 
In this limit we can also obtain the
results by first diagonalizing the matrix $M$ with $F=0$ and then obtaining leading
order corrections in $F$. From the exact formulas we find that this expansion
is valid only in the range,  
\begin{equation}
\delta_F = {x\over y^2} = {F|A-B|\over T^2} <<1
\end{equation}
It is clear that the range over which the leading order results are valid is
considerably limited due to the nonlinear dependence on $T$ in this inequality.

The three eigenvalues, accurate to order $(F/T)^2$ are given by
\begin{eqnarray}
\lambda_1 &=& A - (A-B)\left({F\over T}\right)^2\nonumber\\
\lambda_2 &=& A + {T^2\over (A-B)} + (A-B)\left({F\over T}\right)^2\nonumber\\
\lambda_3 &=& B- {T^2\over (A-B)}
\end{eqnarray}
We display the eigenvectors only to order $\delta_F$ or upto terms linear 
in $F$, since these give the dominant corrections.
The corresponding eigenvectors are given by
\begin{eqnarray}
|\lambda_1> &=& \left(\matrix{1\cr i\delta_F \cr 0}\right)\nonumber\\
|\lambda_2> &=&{1\over D_T} \left(\matrix{\delta_F\cr -i \cr \delta_T}\right)  \nonumber\\
|\lambda_3> &=& {1\over D_T}\left(\matrix{0\cr  \delta_T \cr -i}\right)  
\end{eqnarray}
where $D_T = \sqrt{1+\delta_T^2}$, $\delta_T = T/(A-B)$.
The density matrix elements, accurate to order $\delta_F$ are 
\begin{eqnarray}
\rho_{11}(Z) &=& 
\rho_{11}(0) + \left(i\rho_{12}(0)\delta_F\left[1-e^{i(\lambda_1
-\lambda_2)Z/(2\omega)}\right] + c.c.\right)\nonumber\\
\rho_{22}(Z) &=& {\rho_{22}(0)\over D_T^4} \Bigg|1+\delta_T^2 e^{i(\lambda_2
-\lambda_3)Z/(2\omega)}\Bigg|^2+ \left(i\rho_{12}(0)\delta_F
\left[e^{i(\lambda_2 -\lambda_1)Z/(2\omega)}-1\right] + c.c.\right)\nonumber\\
\rho_{12}(Z) &=& -i\delta_F\rho_{11}(0)\left[1-e^{i(\lambda_2 -\lambda_1)Z/(2\omega)}\right] +i\delta_F\rho_{22}(0)\left[1- e^{i(\lambda_2 -\lambda_1)Z/(2\omega)}\right]\nonumber\\
&+& {\rho_{12}(0)\over D_T^2}\left[ e^{i(\lambda_2 -\lambda_1)Z/(2\omega)}
+ \delta_T^2 e^{i(\lambda_3 -\lambda_1)Z/(2\omega)}\right]
\end{eqnarray}

\subsection{Limit 3:  $F/|A-B|\sim T^2/|A-B|^2<<1$}
We finally consider the case where $x$ and $y^2$ are of the same order but
both are much smaller than unity. The eigenvalues $\bar\lambda_i$ at leading order are
\begin{eqnarray}
 \bar\lambda_1 &=& \sqrt{x^2+y^4/4} - y^2/2\nonumber\\
 \bar\lambda_2 &=& -\sqrt{x^2+y^4/4} - y^2/2\nonumber\\
 \bar\lambda_3 &=& 1+  y^2
\label{lmto}
\end{eqnarray}
with $\lambda_i$ related to $\bar\lambda_i$ by Eq. \ref{eq:lambdabar}. 
The corresponding eigenvectors, accurate to order $y$, are
\begin{eqnarray}
|\bar\lambda_1> &=& {1\over d_1}
\left(\matrix{-1\cr i\bar\lambda_1/x \cr 
y\bar\lambda_1/x}\right)\nonumber\\
|\bar\lambda_2> &=& {1\over d_2}
\left(\matrix{-1\cr i\bar\lambda_2/x \cr 
y\bar\lambda_2/x}\right)\nonumber\\
|\bar\lambda_3> &=& {1\over d_3}\left(\matrix{0\cr  -iy \cr 1}\right)  
\label{lmts}
\end{eqnarray}
where $d_1 = \sqrt{1+\bar\lambda_1^2/x^2}$, and 
$d_2=\sqrt{1+\bar\lambda_2^2/x^2}$ and $d_3 = 1$.
The density matrix elements in this case are given by
\begin{eqnarray}
\rho_{11}(Z) & = & \rho_{11}(0)\left(1-{2\over d_1^2d_2^2}\left[1-\cos(
\Delta_{12}Z)\right]\right) + \rho_{22}(0){2\over d_1^2 d_2^2}\left[
1-\cos(\Delta_{12}Z)\right]\nonumber\\
&-&\Bigg\lbrace\left[i\left({\bar\lambda_1\over xd_1^4}+
{\bar\lambda_2\over xd_2^4} \right) - i{y^2\over xd_1^2d_2^2}
\cos(\Delta_{12}Z) - {2\sqrt{x^2+y^4/4}\over xd_1^2d_2^2}
\sin(\Delta_{12}Z)\right]\rho_{12}(0)
+ c.c.\Bigg\rbrace\nonumber\\
\rho_{22}(Z) & = & \rho_{11}(0){2\over d_1^2 d_2^2}\left[1-\cos(\Delta_{12}Z)\right]+
\rho_{22}(0)\left(1-{2\over d_1^2d_2^2}\left[1-\cos(\Delta_{12}Z)\right]\right) 
\nonumber\\
&+&\Bigg\lbrace -i{\bar\lambda_1\over xd_1^2}\left[1-e^{i\Delta_{12}Z}\right]
\left[\left({\bar\lambda_1\over xd_1}\right)^2 + \left({\bar\lambda_2\over xd_2}\right)^2 e^{-i\Delta_{12}Z} \right]\rho_{12}(0)
+ c.c.\Bigg\rbrace\nonumber\\
\rho_{12}(Z) &=& i{\bar\lambda_1\over xd_1^2}\left[1-e^{-i\Delta_{12}Z}\right]
\left[{1\over d_1^2} + {1\over d_2^2} e^{i\Delta_{12}Z} \right]\rho_{11}(0)
\nonumber\\
&+&i{\bar\lambda_1\over xd_1^2}\left[1-e^{i\Delta_{12}Z}\right]
\left[\left({\bar\lambda_1\over xd_1}\right)^2 + \left({\bar\lambda_2\over xd_2}\right)^2 e^{-i\Delta_{12}Z} \right]\rho_{22}(0)
\nonumber\\
&+&\left[{1\over d_1^2} + {1\over d_2^2}e^{i\Delta_{12}Z}\right]
\left[\left({\bar\lambda_1\over xd_1}\right)^2 + 
\left({\bar\lambda_2\over xd_2}\right)^2 e^{-i\Delta_{12}Z}\right]\rho_{12}(0)
\nonumber\\
&-&{2\over d_1^2 d_2^2}[1-\cos(\Delta_{12}Z)]\rho_{12}^*(0)
\end{eqnarray}
where $\Delta_{12} = (\lambda_1-\lambda_2)/(2\omega)$. 
We point out that in the density matrix elements we have kept only the leading
order terms. The terms of order $y^2$ have been dropped, unless they come 
multiplied by the distance factor $Z$. This is reasonable since for very 
large $Z$ values such terms may give significant contribution but are 
otherwise expected to be negligible compared to the terms we have kept. 

\section{Contribution of the Longitudinal Component}
In the full mixing matrix (\ref{mf-1}) we have so far
ignored the longitudinal mixing 
terms, i.e., $M_{34}$ and $M_{43}$. We now determine the contribution of 
the longitudinal mode by treating these terms as perturbation. 
The matrix element $M_{34}$ 
of the mixing matrix (\ref{mf-1}) is given by,
\begin{equation}
|M_{34}| = N_Lb_{\mu}^{(2)}\tilde{u}^{\mu} = {\omega_p\over2M}|\vec B|
|\sin\theta|\left(1-{\omega_p^2\over\omega^2}\right)^{1\over2}
\end{equation}
%\begin{equation}
%|M_{24}| = {|\vec B|\omega|\sin\theta|\over M}
%\end{equation}
The full
$4\times4$ matrix, given in Eq. \ref{mf-1}, may be expressed as, 
\begin{eqnarray}
M &=& M_0 + M^{'} \nonumber \\
	&=& \left[\matrix{A & iF & 0 & 0 \cr
		-iF & A & 0 & -iT \cr
		0 & 0 & \Pi_L & 0 \cr
		0 & iT & 0 & 0} \right] + \left[\matrix{ 0 & 0 & 0 & 0 \cr 
						0 & 0 & 0 & 0 \cr
						0 & 0 & 0 & iL \cr
						0 & 0 & -iL & 0}\right]
\end{eqnarray}
where $L \approx -\omega_p|\vec B|\sin\theta/(2M)$ in the limit $\omega>>
\omega_p$. Here we treat $M'$ as a perturbation. 
The contribution of the perturbation is small as long as 
\begin{equation}
{L\over |\lambda_i-\lambda_j|} <<1\,,
\end{equation}
where $\lambda_i$ are the unperturbed eigenvalues. In the limit of high 
frequencies, where the pseudoscalar-photon mixing effect may dominate, this
difference is of order $T^2/A$, ignoring the pseudoscalar mass. One 
can easily check that $LA/T^2<<1$ for a wide range of parameters. In 
the low frequency limit, where Faraday effect dominates, the contribution
of the longitudinal mode is also small. Hence we find that for a wide 
range of parameters we can treat the longitudinal mode perturbatively.

We next explicitly compute the modification of the eigenvectors and eigenvalues
for the case discussed in section 5.3 above. Similar results apply in all cases.
The eigenvectors and eigenvalues for the unperturbed matrix $M_0$ can
be obtained from section 5.3. 
The unperturbed eigenvalues can be expressed as, 
\begin{eqnarray}
 \lambda^0_1 &=& {T^2\over2A} + A - \sqrt{F^2 + {T^4\over4A^2}}\,, \nonumber\\
 \lambda^0_2 &=& {T^2\over2A} + A + \sqrt{F^2 + {T^4\over4A^2}}\,, \nonumber\\
 \lambda^0_4 &=& -{T^2\over A}\,,
\label{ttl} 
\end{eqnarray}
where we have set $B=m^2_a=0$.
Similarly, it follows that, the unperturbed eigenvectors,
\begin{eqnarray}
|\lambda^0_1> &=& {1\over d_1}
\left(\matrix{-1\cr v_1 \cr 0
\cr w_1}\right)\nonumber\\
|\lambda^0_2> &=& {1\over d_2}
\left(\matrix{-1\cr v_2 \cr 0 
\cr w_2}\right)\nonumber\\
|\lambda^0_4> &=& {1\over d_4}\left(\matrix{0\cr v_4 \cr 0 \cr 
w_4}\right)  
\label{ttc}
\end{eqnarray}
where, 
$v_1 =i\bar\lambda_1/x$, $v_2 =i\bar\lambda_2/x$, $v_4=-iy$, 
$w_1=y\bar\lambda_1/x$,  $w_2=y\bar\lambda_2/x$, $w_4=1$, $d_4=1$ and
$d_1$, $d_2$, $\bar\lambda_1$ and $\bar\lambda_2$ are defined in 
section 5.3.
We also have,
\begin{eqnarray}
\lambda^0_3 &=& \Pi_L \nonumber \\
|\lambda^0_3> &=& 
\left(\matrix{0\cr 0 \cr 
1\cr 0}\right)
\end{eqnarray}

The leading order corrections to the eigenvalues are given by,
\begin{equation}
M_{jj}^{'} = \left<\lambda_j^0|M^{'}|\lambda_j^0\right>
\end{equation}
For all values for j, the leading order 
corrections to the eigenvalues due to the 
longitudinal perturbative piece vanish. The corrections to  
the eigenvectors are as follows,
\begin{equation}
C_{jk} = {\left<\lambda_j^0|M^{'}|\lambda_k^0\right>\over\left(\lambda_k - \lambda_j\right)}
\end{equation}
We can verify that if $k\ne 3$ then the corrections 
to the eigenvectors are zero. 
Hence for $j=1,2,4$, we have,
\begin{equation}
|\lambda_j> =   \sqrt{1 - |C_{j3}|^2}\ |\lambda_j^0>+C_{j3}|\lambda_3^0>\ .
\end{equation}
where,
\begin{equation}
C_{j3} = {-iL{w}_j\over  d_j\left(\lambda_3 - \lambda_j\right)} 
\end{equation}
Similarly, for $|\lambda_{3}>$ we can write,
\begin{equation}
|\lambda_3> =  \sqrt{1 - \sum_{j=\lbrace1,2,4\rbrace}|C_{3j}|^2}\ |\lambda_3^0>+\sum_{j=\lbrace1,2,4\rbrace}C_{3j}|\lambda_j^0>
\end{equation}
We find that for a wide range of parameters, $|C_{j3}|=|C_{3j}|<<1$.
Hence we see that the contribution of the longitudinal part is small and
can be neglected at leading order.

%
%
%

%%%%%%%%%%%%%%%%%%%%%%%%%%%%%%%%%%%%%%%
\section{Conclusion}
%%%%%%%%%%%%%%%%%%%%%%%%%%%%%%%%%%%%%%%
%
In this paper we have given a general treatment of pseudoscalar-photon
mixing in a magnetized medium. We solve the resulting coupled wave equations
in several different regimes with the assumption that the frequency
of the electromagnetic wave is much larger than the plasma frequency. 
The problem is a little complicated due to the presence of near degeneracy
in the two transverse modes of the electromagnetic wave.
We find, as expected, that at very high frequencies, the Faraday effect
gives negligible contribution and the standard treatment of pseudoscalar-photon
mixing is applicable. However at low frequencies, Faraday effect cannot 
be neglected. We extend the pseudoscalar-photon oscillation formulas so that
they are applicable in this regime also. We find that in this case the 
oscillation effect is considerably modified due to the presence of near
degeneracy in the mixing matrix.  
These results may be useful in many laboratory experiments and in 
astrophysical observations. 
%
%%%%%%%%%%%%%%%%%%%%%%%%%%%%%%%%%%%%%%%
\section{Acknowledgement}
We thank Amit Banerjee for collaborating in the initial stages of this work.
S. Mandal thanks a Department of Science and Technology (DST) for
financial support. P. Jain thanks John P. Ralston for pointing out potential
problems that can arise in the phenomenon of pseudoscalar-photon mixing
due to the presence of near
degeneracies.
%%%%%%%%%%%%%%%%%%%%%%%%%%%%%%%%%%%%%%%
%

\end{document}